\documentclass[useAMS,usenatbib]{mn2e}

\usepackage{aas_macros,amsmath,graphicx,lineno,longtable,amssymb,lastpage}
\usepackage[version=3]{mhchem}

\newcommand{\rearth}{$R_{\oplus}$}
\newcommand{\mearth}{$M_{\oplus}$}
\newcommand{\water}{H$_2$O}
\newcommand{\kep}{{\it Kepler}}
\newcommand{\logg}{$log~g$ }

\newcommand{\teff}{$T_{\rm eff}$}
\newcommand{\tess}{{\it TESS}}
\newcommand{\cheops}{{\it CHEOPS}}

\newcommand{\plato}{{\it PLATO}}
\newcommand{\feh}{[Fe/H]}
\newcommand{\jwst}{{\it JWST}}
\newcommand{\ariel}{{\it ARIEL}}
\newcommand{\teq}{$T_{eq}$}

\title[Multi-Observatory Transit Photometry]{Exoplanet Characterization by Multi-Observatory Transit Photometry with \tess{} and \cheops{}}

\author[Gaidos et al.]{E. Gaidos,$^{1,2}$\thanks{Visiting Scientist, Center for Space and Habitability, University of Bern, and International Space Science Institute, Bern, Switzerland. E-mail:  gaidos@hawaii.edu (EG)} D. Kitzmann,$^{3}$ K. Heng$^{3}$\\
$^{1}$Department of Geology \& Geophysics, University of Hawaii at M\={a}noa, Honolulu, Hawaii 96822 USA\\
$^{2}$Fulbright Fellow, Institute for Astrophysics, University of Vienna, Vienna, 1180 Austria\\
$^{3}$Center for Space and Habitability, University of Bern, Bern, 3012 Switzerland
}

\begin{document}
\date{Submitted to MNRAS}
\pagerange{\pageref{firstpage}--\pageref{lastpage}} \pubyear{2017}

\maketitle

\label{firstpage}

\begin{abstract}
Space-based photometric surveys have discovered large numbers of planets transiting other stars, but these observe in a single band-pass and yield only the planet radius, orbital period, and transit duration.  Information on the masses, compositions, and any atmospheres of these planets requires additional observations from the ground or space.  The {\it Transiting Exoplanet Survey Satellite} (\tess{}) will yield thousands of planets around bright stars suitable for such follow-up.  In the absence of spectroscopy or spectrophotometry from space, observations through the different pass-bands of multiple space telescopes provide some spectral information useful for identifying false positive signals, distinguishing between reflected light and thermal emission from hot Jupiters, and detecting Rayleigh scattering by planetary atmospheres.   We calculated the expected difference in transit depths measured by the \tess{} and {\it Characterizing Exoplanets Satellites} (\cheops{}) missions, which will be more sensitive to redder and bluer optical wavelengths, respectively.  The difference due to companion or background stars is small ($<3$\% for main sequence companions) and likely to be negligible and undetectable.  For only a few "hot" Jupiters, can combined photometry disambiguate between the reflected and thermal signals from planets.  However, Rayleigh scattering by hazy atmospheres with particles sizes near 0.04 $\mu$m and at pressure altitudes above $\sim 1$~mbar can be detected for $\sim$ 100 transiting planets, assuming every planet has such an atmosphere.  Hazes with this characteristic particle size do not obscure observations at longer (near-infrared) wavelengths; \cheops{} follow-up of \tess{}-detected planets could thus identify candidates suitable for further study with the {\it James Webb Space Telescope}.  
\end{abstract}


\section{Introduction}

Exoplanet research is moving beyond an era dominated by discovery and statistics and into one in which characterization of planets and their atmospheres takes a leading role.  Pioneering surveys by ground-based telescopes using the Doppler radial velocity (RV) method \citep{Udry2007}, and the CoRoT satellite using the transit photometry method \citep{Deeg2013} segued to the \kep{} mission and its transit survey of almost 200,000 stars \citep{Borucki2010}.  The yield of thousands of candidate or confirmed exoplanets has enabled statistical descriptions of the planet population down to Earth-size or smaller.

However, the information provided by a transiting exoplanet mission like \kep{} and its successor {\it K2} is limited to the orbital period and radius of the planet, assuming the radius of the star is known\footnote{The transit duration provides limited information on the orbital elements and/or density of the star}.  This is true of any ``white light" measurement where only the signal integrated over the full wavelength range of the instrument response is recorded.  To validate and characterize transit detections, additional data obtained from the ground or space are usually needed.  Future space-based exoplanet surveys, i.e. the Transiting Exoplanet Survey Satellite \citep[\tess,][]{Ricker2014} and the Planetary Transits and Oscillations \citep[\plato,][]{Rauer2014} will target brighter stars suitable for such follow-up over much of the sky.  In addition to RV observations to measure planet mass, spectral (wavelength-dependent) information is valuable or crucial in at least three other aspects: validation of the planet and confirmation of its host star, distinguishing between reflected and emitted light from the planet as observed during opposition and secondary eclipse, and detection of any atmosphere in transmitted light during a primary transit.  

Validation of the transiting object as a planet and identification of the correct host star often requires other observations, sometimes at multiple wavelengths. To cover large areas of sky, transit surveys use large pixel sizes (21'' for \tess), meaning a non-negligible probability that background stars or stellar companions contribute to the flux in the photometric aperture of a target.  The additional sources dilute the transit signal, meaning that the transiting object is actually larger than apparent.   In addition, a fainter, blended star rather than the target star may be the source of the signal.  Blended eclipsing binaries with a large contrast ratio or at much greater distances than the target star can produce "false positives" that mimic some aspects of a planetary transit signal.   These scenarios have motivated a variety of methods to identify false positives, including analysis of the transit photometry itself \citep{Seager2003,Batalha2010,Bryson2013,Mullally2016}, and high-resolution imaging of the host star to identify other blended sources in the transit survey \citep[e.g.,][]{Everett2015}.  Another approach is to measure the transit depth at different wavelengths; if the host star has a different color than other stars that contribute to the photometric flux then the transit depth will vary with wavelength \citep{Tingley2004,Colon2012,Tingley2014,Desert2015}.  For example, the transit will be deeper at longer wavelengths if the actual source of the signal is redder than the target star.     

Additional observations are invariably required to obtain planet properties other than radius and orbital period; foremost among these are planet mass, which can be measured by RV measurements or the method of transit timing variation \citep{Agol2005,Holman2005}, and presence and composition of any atmosphere.  Estimation of both mass and radius allows a planet to be placed on a mass-radius diagram and compared to interior models.  These comparisons can, for example, identify planets that are almost exclusively rocky-metal, with no or only a thin atmosphere, as well as planets with much lower densities that can be explained only by a thick envelope of  hydrogen and helium which are incondensible and have a low molecular weight.  In intermediate regions of the  mass-radius diagram the models are degenerate, i.e. there are multiple possible bulk compositions, and mass and radius alone are not sufficient \citep{Rogers2010}.  In these cases, identification of a H-He envelope by independent means is valuable.  Information on any atmosphere is also crucial to understanding the formation, chemistry, surface conditions, and potential habitability of a planet.

One approach to detecting and characterizing the atmosphere of an exoplanet is by observations at multiple wavelengths when either the planet passes in front of the host star (primary transit), or when the star occults the planet (secondary eclipse) \citep{Seager2008,Encrenaz2015}.  In the former case, variation in the absorption of the atmosphere due to atomic or molecular lines or scattering cause small changes in the apparent radius of the planet with wavelength.  In the latter, the atmosphere manifests itself by reflection and emission and hence changes in the secondary eclipse depth that are wavelength-dependent.  

The primary \tess{} mission will survey most of the sky and is expected to detect $\gtrsim 2000$ planets \citep{Sullivan2015} around relatively bright stars suitable for such follow-up.  These observations can be performed by ground-based telescopes but variable atmospheric absorption and the finite duration of continuous observations limit accuracy and completeness.  Ground-based spectroscopy on the future generation of extremely large telescopes (ELTs) hold promise for exoplanet characterization, but the time available will greatly limit the task of screening thousands of planets.  Highly accurate spectrophotometric observations can be performed above the atmosphere by the {\it Hubble} Space Telescope ({\it HST}) or the future {\it James Webb Space Telescope} (\jwst{}), but these are even more heavily competed for.  What are needed are means to cheaply and rapidly screen most or all \tess{}-detected planets to identify those objects of particularly interest, e.g. those most likely to have a detectable atmospheric signal.  

\begin{figure}
 \includegraphics[width=\columnwidth]{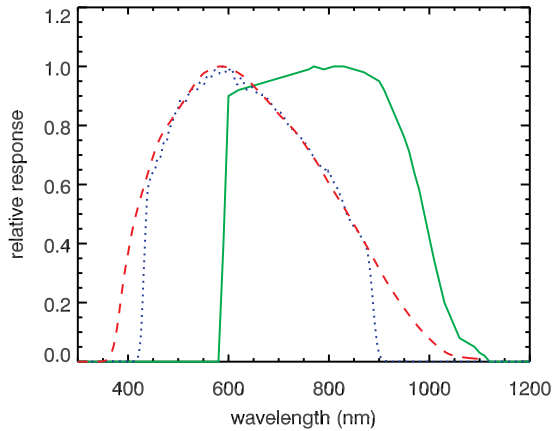}
 \caption{The predicted or established response functions of \tess{} (solid green), \kep{} (dotted blue), and \cheops{} (dashed red), all normalized to unit maximum.}
 \label{fig.bandpass}
\end{figure}

At the time of writing, no space satellite other than {\it HST} offers spectroscopy or multi pass-band photometry from space.   In the absence of such observations, limited spectral information can be still be extracted if the pass-bands of two measurements are substantially different, i.e. if one is redder or bluer than the other.  In particular, the deep-depleted charge-coupled detectors (CCDs) to be used in the cameras of the Transiting Exoplanet Survey Satellite (\tess) have  substantial response at $\lambda = 1\mu$m and beyond \citep{Woods2016}.  This red response distinguishes them from the conventional CCDs used in other cameras, i.e. that of the the NASA \kep{} satellite \citep{Koch2010} and the upcoming Characterizing Exoplanet Satellite \citep[\cheops,][]{Broeg2014} and \plato{} missions, which have declining sensitivity beyond $\lambda = 0.6\mu$m and virtually no response beyond 0.9-1 $\mu$m.  Figure \ref{fig.bandpass} show the relative response functions of the \kep, \cheops, and \tess{} (that of \plato{} is not yet available).     

\citet{Placek2016} demonstrated how the combination of photometry from \kep{} and \tess{} can distinguish between the contributions of reflected light and thermal emission to the total signal from a "hot" Jupiter-size planet on a close-in orbit.   While \kep{} photometry is more sensitive to bluer wavelengths and thus the reflected signal, \tess{} is more sensitive to redder wavelengths and the thermal emission from the hot day-side of the planet.  By observing the depth of the secondary eclipse with both instruments, and making assumptions about the reflected and thermal spectra, the two signals can be separated, allowing both the geometric albedo and the day-side temperature to be independently estimated.  The drawback of this strategy is that the primary \kep{} mission observed only 0.28\% of the sky, and the successor K2 mission is restricted to fields near the plane of the ecliptic, which \tess{} will not observe.\footnote{An extended \tess{} mission could observe the ecliptic.}  

In contrast, \cheops{} can point to most of the sky and is designed expressly for follow-up of known exoplanet systems, including those discovered by \tess{}.  In this work, we show how the combination of measurements by \tess{} and \cheops{} can contribute to characterization of transiting exoplanets and their atmospheres in the three respects described above.   In Sec. \ref{sec.methods} we describe some of the tools and databases used in our calculations.  In Sec. \ref{sec.fp} we analyze the degree to which these measurements can identify the correct host star and identify false positives, i.e. diluted eclipsing binaries.  In Sec. \ref{sec.thermal} we investigate the contribution of \tess{}-\cheops{} measurements to distinguishing between the reflected and thermal emission from planets, i.e. hot Jupiters.  In Sec. \ref{sec.transit} we use model planet atmospheres and empirical or realistic model spectra of stars to determine the magnitude of the \tess{}-\cheops{} signal produced by an atmosphere during a primary transit.  We expand on the last aspect in Sec. \ref{sec.tess} where we consider the detection of atmospheres with different compositions among the entire catalog of simulated \tess{} planets constructed by \citet{Sullivan2015}.  We conclude in Sec. \ref{sec.discussion} with a discussion of systematic effects and future prospects for multi-color transit spectrophotometry both from space and the ground.

\section{Methods}
\label{sec.methods}

{\it Stellar spectra:} Two sets of spectra were used for these calculations.  For K- and M-type dwarf stars (\teff{} $<5400$K), we used actual spectra obtained at visible and near-infrared wavelengths and flux-calibrated by \citet{Mann2015} and Gaidos et al. in prep.  For the full range of spectral types we used model spectra generated by the PHOENIX code in spherical mode by \citet{Husser2013}.  The spectrum of the host star was interpolated from a grid with \teff{}, \logg{}, and [Fe/H].  In cases where the metallicity is not known it was assumed to be solar.

{\it Response functions and calculations:}  The response vs. wavelength of \tess{} was taken from \citet{Ricker2014}. The \cheops{} response function was calculated by multiplying the throughput of the optics by the detector quantum efficiency using data provided by the \cheops{} team (A. Fortier, personal communication).  The \kep{} high spectral resolution response function was constructed by J. van Kleeve and distributed by the NASA \kep{} Guest Observer office\footnote{https://keplergo.arc.nasa.gov/kepler\_response$\_$hires1.txt}.  

{\it Planetary atmosphere transmission model:} For the calculation of the transmission spectra (as a wavelength-dependent apparent planet radius $R_p$) we assumed a reference radius $R_0$ on top of which we place an isothermal atmosphere with a total pressure at $R_0$ of two bar. The pressure is high enough to guarantee that the atmosphere becomes opaque at all wavelengths. The planet is then divided into 200 annuli $p$. The integrated slant optical depth along each tangent height was calculated for via
\begin{equation}
\tau_\lambda(p) = \int_{-\infty}^{\infty} \chi_\lambda(x(p)) \, \mathrm d x
\end{equation}
where $\chi_\lambda$ is the extinction coefficient, including contributions by absorption and scattering, and $x$ the coordinate along the tangent.

The effective tangent height was then calculated via
\begin{equation}
  h_\lambda = \int_0^\infty \left(1 - T_\lambda(p) \right) \mathrm d p
\end{equation}
where
\begin{equation}
  T_\lambda(p) = e^{-\tau_\lambda(p)}
\end{equation}
is the transmission along a given tangent $p$.
Finally, the wavelength-dependent transit radius is given by
\begin{equation}
R_\lambda = R_0 + h_\lambda \ .
\end{equation}  
Additionally, the transmission spectrum was normalised such that within a white-light filter from 0.5 $\mu$m to 1.05 $\mu$m a given planetary radius is obtained.

The absorption coefficients for \ce{H2O}, \ce{CO2}, \ce{CO}, \ce{CH4}, \ce{NH3}, \ce{C2H2}, and \ce{HCN} were calculated with the HELIOS-k opacity calculator \citep{Grimm2015ApJ...808..182G}. For \ce{H2O}, \ce{CO2}, and \ce{CO} we use the HITEMP2010 line lists \citep{Rothman2010JQSRT.111.2139R}, whereas for the rest, HITRAN2012 \citep{Rothman2013JQSRT.130....4R} was employed.  In addition to these molecules, the lines of the alkali metals \ce{Na} and \ce{K} were also included, using the line data from \citet{Draine2011piim.book.....D}.  Collision induced absorption continua description from HITRAN2010 \citep{Rothman2010JQSRT.111.2139R} was employed for \ce{H2}-\ce{H2} and \ce{H2}-\ce{He} collisions. 

The molecular scattering cross sections were derived via the usual Rayleigh scattering equation
\begin{equation}
  \sigma_\mathrm{rayleigh} = \frac{24 \pi^3 \nu^4}{n_\mathrm{ref}^2} \cdot \left(\frac{n(\nu)^2 - 1}{n(\nu)^2 + 2}\right)^2 \cdot K(\nu) \ ,
\end{equation}
where $\nu$ is the wavenumber, $n$ the refractive index, $n_\mathrm{ref}$ a reference particle number density, and $K$ the King factor.
The corresponding data for \ce{H2} is adopted from \citet{Allen2000asqu.book.....C}, for \ce{H2O} from 
\citet{wagner2008international}, \citet{Murphy1977}, \citet{Sneep2005JQSRT..92..293S} and \citet{Thalman2014JQSRT.147..171T} for \ce{CO}, \ce{CH4}, \ce{CO2}, and \ce{He}, respectively. The atmospheric abundances of the molecules are calculated using an equilibrium gas-phase chemistry \citep{Stock2017InPrep}. The chemistry model contains about 550 gas-phase species (including ions) and is based on the semi-analytic approach described in \citet{Gail2014pccd.book.....G}. The solar elemental abundances are taken from \citet{Asplund2009ARA&A..47..481A}.

For the high-altitude condensates we follow the approach of \citet{Howe2012ApJ...756..176H} and assume a photochemical haze composed of tholin particles. The corresponding refractive index is adopted from \citet{Khare1984Icar...60..127K}. Unless otherwise stated, all hazes are modelled with a monodisperse size distribution and a particle number density of 1000 cm$^{-3}$.

\section{Identification of Host Star in Multiple Systems and False Positives}
\label{sec.fp}

The large angular size of the pixels and photometric apertures of wide-field surveys such as \tess{} means that blends with stars other than the target star will be frequent, and many stars are unresolved multiples, thus it is often necessary to distinguish between multiple possible host stars for the transiting object.  Moreover, a common "false positive" scenarios in wide-field exoplanet surveys is a fainter background or companion eclipsing binary that contaminates the aperture \citep{Brown2003,Evans2010}.  If the colors of the stars that contribute flux to the total measured signal are different, then the transit depth will vary with wavelength and thus the pass-band because only one of the stars is obscured.   The detection of a difference in transit depth between pass-bands thus can indicate a false positive and the {\it sign} of the difference can be used to tell which star is the host.  The fractional change in the transit depth $\delta$ of a planet orbiting star ``1'' due a second star (``2'') is
\begin{equation}
\label{eqn.fp}
\Delta \delta = \int{}\frac{f_{1,A}f_{2,B}-f_{1,B}f_{2,A}}{\left(f_{1,A} + f_{2,A}\right)\left(f_{1,B} + f_{2,B}\right)} d\lambda,
\end{equation}
where $f(\lambda)$ is the signal from source 1 or 2 in pass-band A or B. 

\begin{figure}
 \includegraphics[width=\columnwidth]{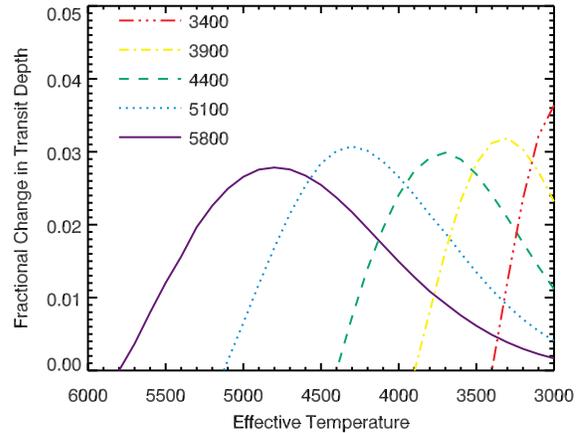}
 \caption{Fractional difference in \cheops{} and \tess{} transit depths due to the presence of an unresolved companion star as a function of the secondary star's \teff{}, for several values of the primary star's \teff{}.}
 \label{fig.fp}
\end{figure}

Using equation \eqref{eqn.fp}, and the spectra described in Sec. \ref{sec.methods}, we calculated the fractional difference in \cheops{} and \tess{} transit depths produced by physical companion, for the case where the stars are unresolved and both on the main sequence and the secondary is cooler and less luminous (Fig. \ref{fig.fp}).  Since equation \eqref{eqn.fp} is an anti-symmetric function for $1 \Leftrightarrow 2$, the sign of the difference will depend on whether the planet orbits the first or second stars.  In all cases, the estimated difference is $<3$\%.  The difference is small for nearly equal brightness binaries because the spectra are similar, whereas much cooler companions will have equal brightness but contribute little to the overall flux.  If the stars are resolved and the secondary contributes only part of its flux to the photometric aperture, the difference will be smaller.

We also considered the case of unrelated background stars.  There are many possible configurations of candidate host stars and background sources, so to generate realistic scenarios, we used the synthetic catalog of \tess{}-detected transiting planets and their host stars generated by \citet{Sullivan2015}.  We constructed a synthetic background star population covering one square degree to $V = 30$ using TRILEGAL v. 1.6 \citep{Girardi2016} with the default settings and "typical" Galactic coordinates $l = 90$ and $b = 30$.  There are negligible numbers of fainter ($V > 30$) stars for these settings.  TRILEGAL outputs the fundamental parameters for each star plus the line-of-sight extinction $A_V$.  The surface density of stars is 10.9 per sq. arc min., thus there is an average of 1.2 background star per 21 arc-sec square \tess{} pixel.

We considered scenarios with a single background star per exoplanet host and simulated 10 such scenarios per host star, randomly drawing the background stars from the synthetic population.  The spectra of the host star and background star were approximated by the PHOENIX spectrum with the closest values of \teff{}, \logg, and \feh, and reddened using the simulated $A_V$ values converted to $E_{B-V}$ assuming a reddening ratio of 3.1 and the reddening curve of \citet{Cardelli1989}.  

\begin{figure}
 \includegraphics[width=\columnwidth]{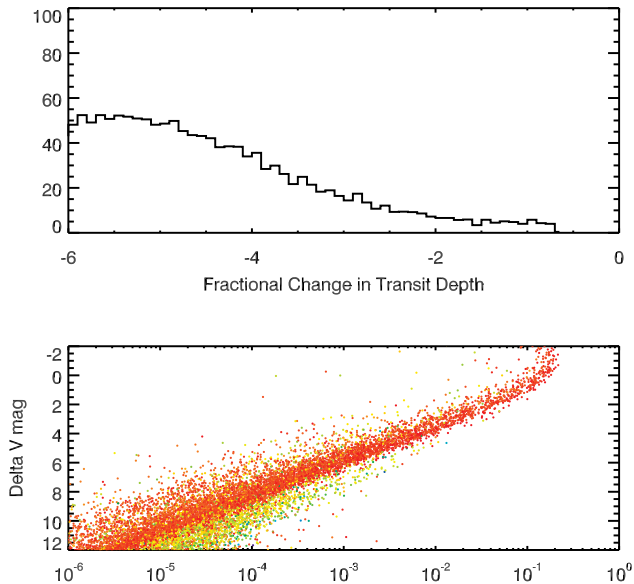}
 \caption{Top panel: Distribution of fractional change in 1984 transit detections simulated by \citet{Sullivan2015}  produced by a single star randomly selected from a background stellar population synthesized by TRILEGAL for $l = 90$, $b=30$.  Bottom panel: Contrast ratio in $V$-band between host and background star vs. transit depth difference.  Points are color-coded according to the \teff{} of the host star, with red points representing cooler stars.}
 \label{fig.fp2}
\end{figure}

We calculated the fractional difference in transit depth using equation \eqref{eqn.fp} and plot this vs. the contrast ratio in magnitudes $\Delta V$ in Fig. \ref{fig.fp2}.  There is the expected correlation where brighter background stars tend to produce larger deviations in transit depth.  However, in only a small fraction of instances (66, 3.4\%) is the difference appreciable ($>$1\%).  In the vast majority of cases, the interloping background star can be clearly identified by ground-based seeing-limited observations ($\sim$1" resolution).  For these cases, the spectrum of the background star can be estimated base on photometric colors and hence the sign and magnitude of the expected difference in transit depth can be calculated and compared with any observed difference.

\section{Disambiguation of Reflected and Emitted Radiation}
\label{sec.thermal}

A sufficiently hot planet emits significant radiation at visible wavelengths \citep{Heng2013}.  \citet{Placek2016} demonstrated how combined observations by \kep{}/K2 and \tess{} during a secondary eclipse by the star can be used to separate the thermal emission and reflected light from a "hot" Jupiter.  The total signal from the planet, as measured by the eclipse depth, is equal to its reflected light plus the Wien tail of the thermal emission, and for bodies with equilibrium temperature $T_{eq} \gtrsim 2000$K the latter is significant.  The thermal component is sensitive to detector sensitivity at the reddest wavelengths and thus is larger for \tess{} than \kep{}.  We performed a similar calculation, comparing the transit depths measured by \cheops{} and \tess{}, the former satellite having a response function similar to that of \kep{} (Fig. \ref{fig.bandpass}).  

\begin{figure}
 \includegraphics[width=\columnwidth]{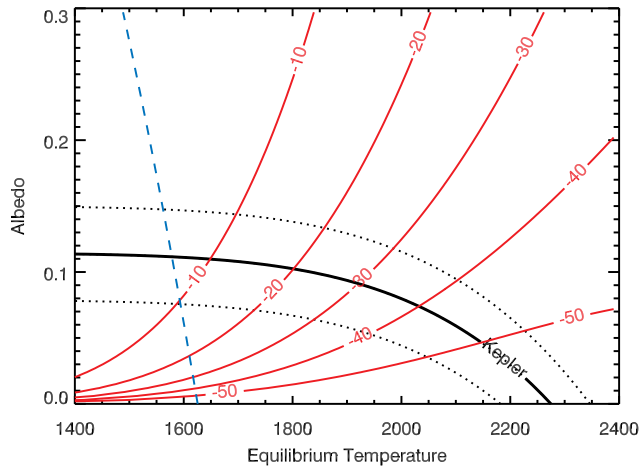}
 \caption{Possible albedo $a$ vs. equilibrium temperature \teq{} for the hot Jupiter \kep{}-43b, based on the \kep{} secondary eclipse depth, assuming completely efficient heat redistribution (solid black line, with $\pm1\sigma$ as dotted black lines).   Theoretical possible values are represented by the blue dashed line.  The secondary eclipse depth measured by \kep{} constrains the combination of \teq{} and albedo $a$ but there is degeneracy between the values.  The red contours are the predicted \% difference in secondary eclipse depth between \cheops{} and \tess{} observations, showing how a measurement can break the degeneracy.}
 \label{fig.albedo}
\end{figure}

Figure \ref{fig.albedo} demonstrates the principle of this analysis with the hot Jupiter \kep{}-43b.  This system was selected because \citet{Angerhausen2015} finds a negative and obviously nonphysical albedo for the planet.  The properties of the star were taken from \citet{Bonomo2015} and secondary eclipse parameters from \citet{Angerhausen2015}.  The calculation assumes black-body thermal emission and efficient redistribution of heat around the planet.  The constraint from the \kep{}-measured secondary eclipse depth is plotted as the solid black contour in Fig. \ref{fig.albedo} with the dashed lines representing $\pm 1\sigma$.  Contours of constant \% difference between the predicted depth of the secondary eclipse in the \cheops{} vs. the \tess{} band-passes are plotted as red lines.  The dashed blue line is the value of \teq{} assuming efficient heat redistribution.  If heat redistribution is efficient, the difference between \cheops{} and \tess{} should be $\lesssim 10$\%.  A deeper secondary eclipse signal in the \tess{} pass-band would indicate a higher day-side temperature and less efficient heat transport to the night side.       

\begin{figure}
 \includegraphics[width=\columnwidth]{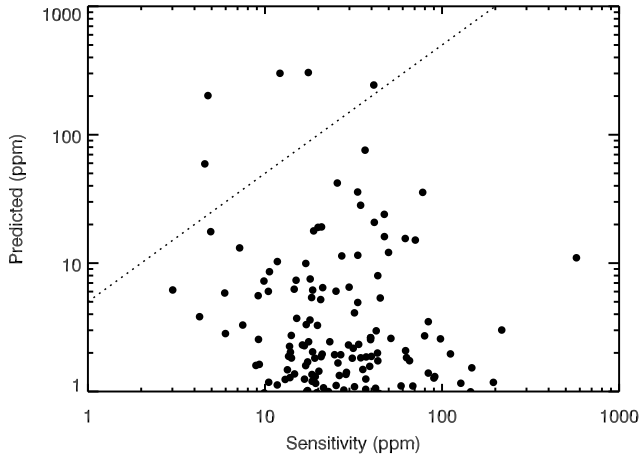}
 \caption{Predicted difference in secondary eclipse depth between \cheops{} and \tess{} in ppm vs. the expected detection sensitivity of \tess{}, for the simulated catalog of \tess{} detections by \citet{Sullivan2015}.  An albedo of 0.1, black-body thermal emission, and perfect redistribution of heat around each planet are assumed.  The dotted line is a detection threshold of five times the sensitivity, and simulated planets above the line are expected to have detectable differences.  Most \tess{} planets have predicted to exhibit differences of $<1$ ppm and are not shown.}
 \label{fig.albedo2}
\end{figure}

We calculated the \cheops{}-\tess{} difference in the secondary eclipse depth for the simulated planets in the \citet{Sullivan2015} catalog, adopting a representative value of 0.1 for the albedo \citep{Shporer2015}, blackbody emission, and perfect redistribution of heat.  Since the duration of the primary transit and secondary eclipse are equal for circular orbits, the photometric sensitivity to the two signals are equal and obtained by dividing the transit depth by the SNR estimated by \citet{Sullivan2015}.  Figure \ref{fig.albedo2} plots the expected \cheops{}-\tess{} difference vs. the expected \tess{} sensitivity.  The dotted line is a $5\sigma$ detection threshold, assuming that \tess{} observations limit the sensitivity\footnote{The SNR of a {\it single} \cheops{} transit observation is expected to be similar to that of the total \tess{} photometry.}  Only five planets out of nearly 2000, all hot Jupiters, are expected to have a detectable \cheops{}-\tess{} difference.  This number will be larger if heat is not efficiently redistributed and the planets have hotter day-sides.  In Appendix \ref{sec.phase} we also consider the potential for detection of differences in the relative {\it phase} in the peak of the light curve due to the non-uniform distribution of clouds on the day-side and transport of heat to the night-side.

\section{Transmission Spectra of Planets}
\label{sec.transit}

It is possible to detect the atmosphere of a planet by comparing spectra obtained during and before/after a transit.  The atmosphere manifests itself as a wavelength dependence in the transit depth and thus the apparent radius of the planet.  The planet will appear larger at  wavelengths where constituents in its atmosphere absorb or scatter some of the light of the host star.  Atoms and molecules will absorb at discrete wavelengths whereas Rayleigh scattering by these and particles in hazes/clouds can produce a larger apparent radius at bluer wavelengths and a negative spectral slope.  This wavelength dependence, integrated over the total response function of individual observatories, can produce differences in apparent transit depth.  The magnitude of this effect depends on the scale height of the atmosphere $h$ relative to the radius of the planet $R_p$, the composition of the atmosphere, and the spectrum of the host star, which effectively ``weights'' different wavelengths.

We calculated the difference in transit depth between \cheops{} and \tess{} for the reference case of a GJ~1214b-like planet ($R_p = 2.6$\rearth, $M_p = 6$\mearth) but with a solar-composition atmosphere (mean molecular weight $\bar{\mu} = 2.33$), and a \teq{} of 1000~K.  The ratio $h/R_p$ is 0.023.  We considered several different atmosphere scenarios with solar composition: two "clear" atmospheres with no hazes or clouds, with and without \water-vapor, and six hazy atmospheres with characteristic particle sizes $r_{\rm part}$ of 0.01, 0.05, and 0.1 $\mu$m, and cloud/haze-top pressure heights $z_{\rm cloud}$ of 1 $\mu$bar and 1 mbar.  We considered a range of host star spectra, i.e. \teff{} of 3000-7000K with solar metallicity and main-sequence surface gravity as proscribed by the empirical $M_*$-$R_*$ relation of \citet{Pecault2013}.  

\begin{figure}
 \includegraphics[width=\columnwidth]{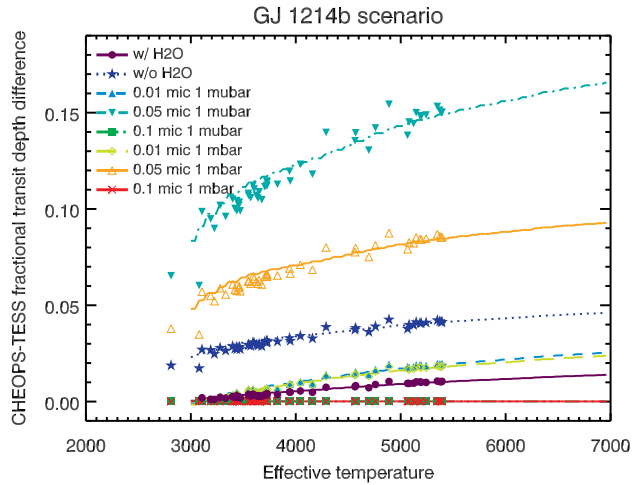}
 \caption{Predicted fractional difference in transit depth between \cheops{} and \tess{} for a planet with the surface gravity of GJ~1214b and an equilibrium temperature of 1000K.  Eight different atmosphere scenarios are considered; two clear atmosphere with and without water, and six hazy atmosphere with different particle size and haze heights.  The curves are based on PHOENIX model spectra and the points are based on actual flux-calibrated spectra of K and M dwarfs.}
 \label{fig.gj1214b}
\end{figure}

The fractional transit depth difference is plotted vs. \teff{} in Fig. \ref{fig.gj1214b}.  In all cases with \teff{} $>3000$K the \cheops{} transits are deeper (the difference is positive) because a planet's atmosphere is usually more opaque and/or scattering at shorter wavelengths, where \cheops{} is more sensitive.  The difference increases with increasing stellar \teff{} because hotter stars have more flux at $\lambda < 580$~nm where \cheops{} but not \tess{} is sensitive.  The calculations using empirical spectra (points in Fig. \ref{fig.gj1214b}) confirm the calculations based on models of stellar spectra and show that the variations due to stellar metallicity and surface gravity is modest.  

Transits of planets with clear atmospheres lacking clouds or hazes will appear similar ($<$ 5\% difference) in observations by \tess{} and \cheops{}.  This is particularly true of clear atmospheres with \water{} vapor, which heavily absorbs at the reddest wavelengths and balances molecular Rayleigh scattering at shorter wavelengths.  Atmospheres with aerosols dominated either by $0.01 \mu$m or $0.1 \mu$m particles also do not produce much of a difference in transit depth.  Atmospheres containing a haze with a 0.05 $\mu$m particles generate the largest differences (up to $\approx 15$\%), with the higher haze (1 $\mu$bar pressure height) producing the largest signal.  Aerosols of this size are efficient Rayleigh-scatterers at wavelengths probed best by \cheops{} but inefficient at the redder wavelengths sensed by \tess{}.   

\begin{figure}
 \includegraphics[width=\columnwidth]{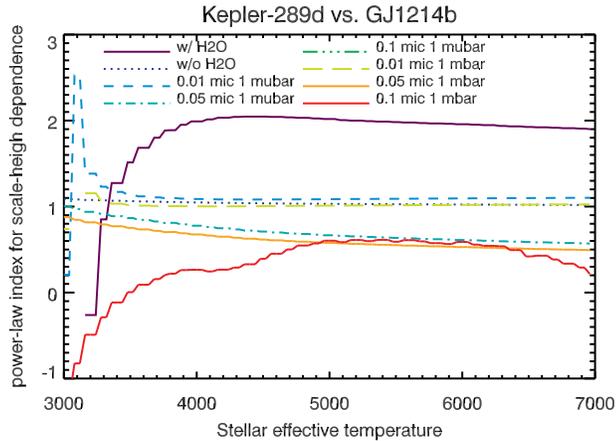}
 \caption{Power-law index $\beta$ relating the predicted fractional difference in transit depths observed by \cheops{} and \tess{} to the ratio of scale height to planet radius $h/R_p$.  These use the calculations for planets with the masses and radii of \kep{}-289d and GJ~1214b, but with solar-metallicity atmospheres.  $\beta = 1$ is the linear dependence predicted for a simple scattering/absorbing atmosphere.  See text for an explanation of deviations from nonlinearity and dependence on stellar \teff{}.}
 \label{fig.scaleheight}
\end{figure}

In principle, the fractional contribution of a homogeneous absorbing/scattering atmosphere to the transit depth should be proportional to $h/R_p$, thus the fractional difference in transit depth due to an atmosphere should also scale linearly with that ratio.  We tested this assumption by performing the same calculation for the Earth-size planet \kep{}-289d, which has an estimated $h/R_p$ larger by a factor of 1.58 than GJ~1214b (assuming identical \teq{} and atmospheric composition).  We used the GJ~1214b and \kep{}-289d results to calculate a power-law index $\beta$ which equates the ratio of $h/R_p$ values to the ratio of \cheops{}-\tess{} difference $\Delta$, i.e. $\Delta \propto (h/R_p)^{\beta}$.  $\beta = 1$ represents linearity.  Figure \ref{fig.scaleheight} plots $\beta$ as a function of stellar \teff{} for the eight different scenarios.  The reader is reminded that this is the {\it scale-height dependence} of the difference in transit depth, not the difference itself.

Several scenarios follow approximate linearity ($\beta \approx 1$).  These include the clear, \water{}-free atmosphere in which Rayleigh scattering dominates at visible wavelengths.  Hazy atmospheres containing particles with $r_{\rm part} = 0.01 \mu$m, which act as Rayleigh-scatterers at all wavelengths, have a similar behavior.  Intermediate-size particles ($r_{\rm part} = 0.05 \mu$m) act as Rayleigh scatterers only at the shorter wavelengths, but produce a flat spectrum at longer wavelengths.  This introduces a spectral dependence on the signal and a slow change with \teff{}.   In the atmosphere with large ($r_{\rm part} = 0.1 \mu$m) haze particles at high altitude ($z_{\rm cloud} = 1 \mu$bar), all spectral features are obscured.  There is no difference between \cheops{} and \tess{} observations because the spectrum is flat at all wavelengths and the scale-height dependence is ill-defined.  In an atmosphere with particles confined to a lower pressure altitude ($z_{\rm cloud} = 1$~mbar), the only atmospheric features are the resonant absorption lines of neutral sodium and potassium doublets near 589 and 768~nm; these produce some difference in the transit depth.  The individual contributions of the lines at different wavelengths also makes the signal dependent on the stellar spectrum and hence \teff{}.

\section{Detecting the Hazy Atmospheres of \tess{} Planets}
\label{sec.tess}

The possibility that planets with hazy atmospheres can exhibit significantly different transit depths in \cheops{} and \tess{} observations motivated an estimate of the total number of \tess{}-detected planets with detectable differences.   We calculated the transit depths that \cheops{} and \tess{} would observe for the planets in the simulated detection catalog of \citet{Sullivan2015}.  

The spectral difference will depend on $R_p$, $h$, $T_eq$, $r_{\rm part}$, $z_{\rm cloud}$, the composition of the atmosphere, and the stellar spectrum (i.e., \teff{}).  For a homogeneous scattering or absorbing atmosphere, the fractional difference in transit depth due to the atmosphere scales as $h/R_p$.  This assumption clearly fails for certain atmospheres, e.g. atmospheres with large hazes sizes, hazes at low altitude and clear atmospheres containing \water{}-vapor (Fig. \ref{fig.scaleheight}).  But in such cases the difference in transit depth is small and probably not detectable, regardless.
 
\begin{figure}
 \includegraphics[width=\columnwidth]{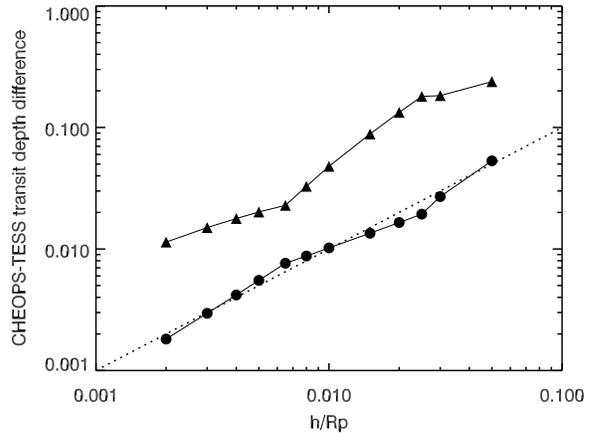}
 \caption{Fractional difference in \cheops{}-\tess{} transit depth vs. atmospheric scale-height to planet radius ratio $h/R_p$ for planets having solar-metallicity atmospheres with hazes with $z_{\rm cloud}  1 \mu$bar pressure height.  The top and bottom curves are for $r_{\rm part} = 0.05$ and 0.01 $\mu$m hazes, respectively.}
 \label{fig.hr}
\end{figure}

We performed a set of calculations with different values of $h/R_p$ for high-altitude haze particles to demonstrate the approximate linear dependence for the most detectable cases.  If this scaling holds one can estimate the signal from any planet by calculating $h/R_p$ and scaling the fractional transit depth difference determined for a reference case, i.e. a GJ~1214b-like planet.  Figure \ref{fig.hr} shows the case for $r_{\rm part} = 0.01$ and 0.05 $\mu$m, all for $z_{\rm cloud} = 1 \mu$bar, solar-metallicity composition, and $T_{eq} =1000$K.  The best-fit power-law indices are 0.96 and 1.09, respectively, demonstrating linearity.

\begin{figure}
 \includegraphics[width=\columnwidth]{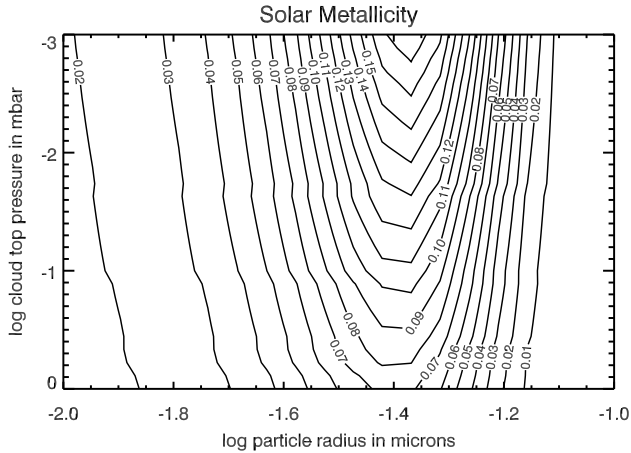}
 \caption{Fractional difference in \cheops{}-\tess{} transit depth vs. haze parameters for a planet with $T_{eq} = 1000$K, $h/R_p = 0.023$, a solar-metallicity atmosphere ($\bar{\mu} = 2.33$), and a solar-type spectrum for the host star.}
 \label{fig.grid1solar}
\end{figure}

\begin{figure}
 \includegraphics[width=\columnwidth]{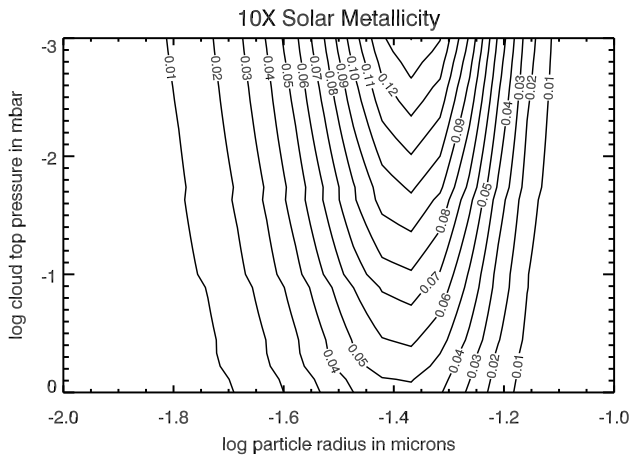}
 \caption{Same as Fig. \ref{fig.grid1solar} but for planet atmospheres with $10 \times$ solar metallicity ($\bar{\mu} = 2.56$).}
 \label{fig.grid10solar}
\end{figure}

We next calculated the \cheops{}-\tess{} difference in transit depth for the reference case (GJ~1214b, 1000K) and ranges of haze particle sizes and pressure altitudes.  Figures \ref{fig.grid1solar} and \ref{fig.grid10solar} show the results for a solar-metallicity composition ($\mu = 2.33$) and $10\times$ solar-metallicity composition, using a solar-type host star.  These show the expected sensitivity to particle size, with $\sim 0.04\mu$m particles producing the maximum signal, as well as the dependence on pressure height, with higher hazes intercepting more starlight and producing a greater difference.  Also as expected, more metal-rich atmospheres have a smaller scale height and generate a smaller signal.  Since Rayleigh scattering occurs only at wavelengths less than the particle size, there is also a dependence on the host star spectrum:  the particle size producing the maximum difference is longer for planets transiting cooler stars which emit more at longer wavelengths.  

The number of simulated \tess-detected planets with a detectable difference in \cheops{} and \tess{} transit depths was calculated for a range of haze particle sizes and pressure heights.  Detection was assumed to require SNR $>5$, with the \tess{} signal-to-noise calculated by \citet{Sullivan2015} and unlimited \cheops{} signal-to-noise.  The scale height for each planet was calculated assuming the radius given by \citet{Sullivan2015}, a mass from the mass-radius relation of \cite{Chen2017}, \teq{} was calculated using the irradiance level given by \citet{Sullivan2015} and assuming an albedo of 0.1.  The difference in transit depths was calculated using a grid of transit depth differences with particle size and cloud-top height calculated for the nominal GJ~1214b-like case with \teq{} = 1000K and solar metallicity atmosphere ($h/R = 0.023$), and assuming linear scaling with $h/R$ as demonstrated in Fig. \ref{fig.hr}.

\begin{figure}
 \includegraphics[width=\columnwidth]{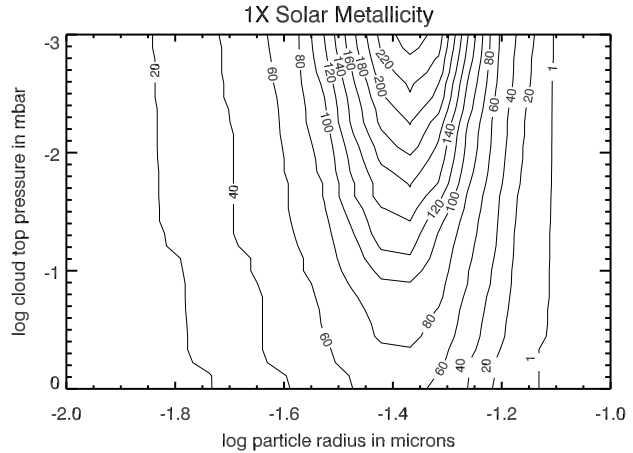}
 \caption{Number of significant ($5\sigma$) detections of a difference between \cheops{}- and \tess{}-measured transit depths for the simulated catalog of \tess{} planet detections by \citet{Sullivan2015}, assuming every planet has a hazy atmosphere, vs. characteristic haze particle size and haze (cloud-top) altitude.  A solar metallicity atmosphere is assumed.}
 \label{fig.yield1solar}
\end{figure}

Figure \ref{fig.yield1solar} plots the numbers of atmosphere detections vs. haze parameters, assuming every planet has such an atmosphere.  The distribution reflects the magnitude of the signal in Fig. \ref{fig.grid1solar} and peaks at 262 detections, or about 13\% of the total.  The overall yield is sensitive to the heavy element abundance of the atmosphere via its influence on the scale height, but the trend is not monotonic because of the effect on atmospheric opacity as well.  In the case of atmospheres with $10\times$ solar metallicity ($\bar{\mu} = 1.56$), the distribution is similar, but the numbers are smaller (Fig. \ref{fig.grid10solar}) and peak at 155.  For $30\times$ ($\bar{\mu}=3.02$) and $50\times$ ($\bar{\mu} = 3.48$) solar metallicity, the distributions (not shown) peak at 179 and 143, respectively.  Note that these last three numbers do not differ by amounts much larger than the expected variance due to counting statistics.

\begin{figure}
 \includegraphics[width=\columnwidth]{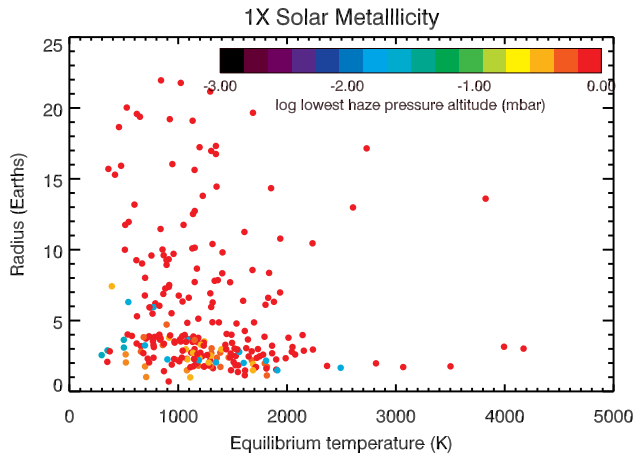}
 \caption{Equilibrium temperature and radius of planets with significant detections of a difference in the \cheops{} and \tess{} transit depths from the \citet{Sullivan2015} for any of the haze scenarios considered in Fig. \ref{fig.yield1solar}.  The color of the point is coded to the lowest pressure altitude (largest pressure) of the cloud-top at which the haze could still be detected.}
 \label{fig.detect1solar}
\end{figure}

\begin{figure}
 \includegraphics[width=\columnwidth]{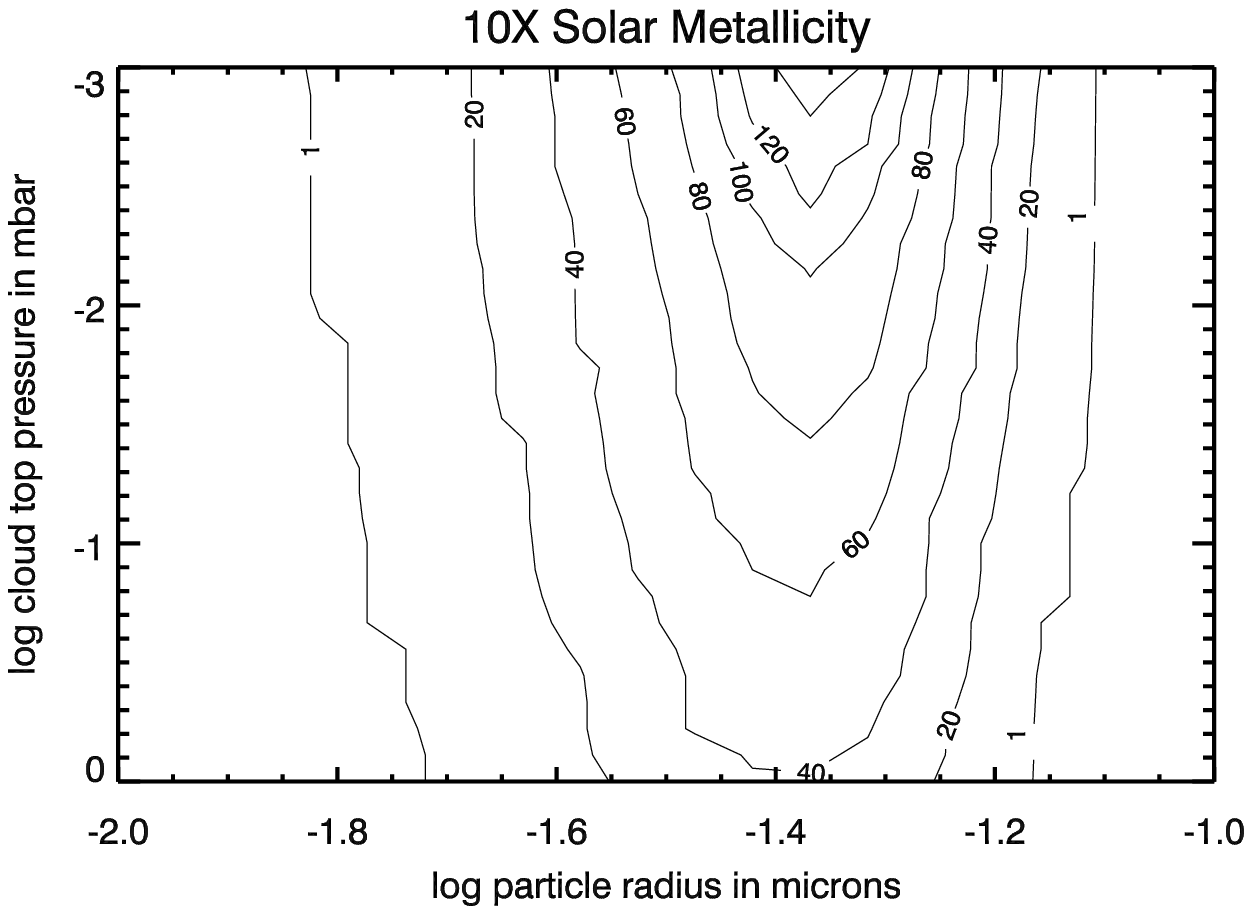}
 \caption{Same as Fig. \ref{fig.yield1solar}, except for atmospheres with $10\times$ solar metallicity.}
 \label{fig.yield10solar}
\end{figure}

\begin{figure}
 \includegraphics[width=\columnwidth]{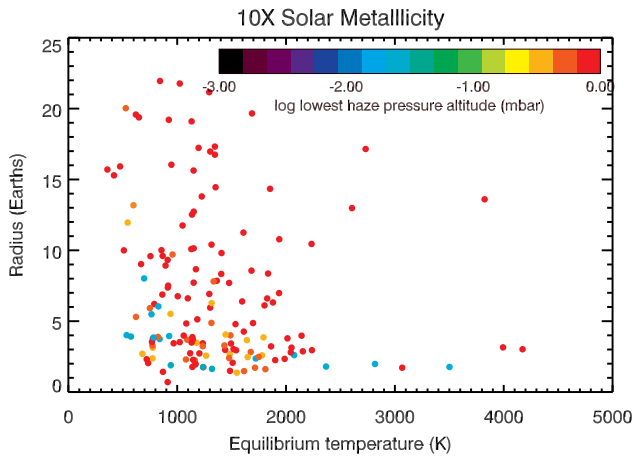}
 \caption{Same as Fig. \ref{fig.detect1solar}, except for atmospheres with $10\times$ solar metallicity.}
 \label{fig.detect10solar}
\end{figure}

\section{Summary and Discussion}
\label{sec.discussion}

We have examined three ways in which differences in the response function of \cheops{} and \tess{} will manifest themselves.  First, a blended companion star or unrelated background star with a different \teff{} and spectrum will produce a difference in transit depth with a sign that depends on whether the planet orbits around the principal target star or the companion/background star.  However, this effect is predicted to be both rare and minor.  While unrelated blends are likely for \tess{} (21" pixels), they are very unlikely for \cheops{} (1" pixels).  Regardless, simulations suggest that in only 3.4\% of cases will the effect exceed 1\%.   Companion stars (separations of a few AU or less) that are unresolved both by \tess{} and \cheops{} can give rise to transit depth differences of up to 3\% (maximal when the secondary star is $\sim$800K cooler than the primary).

Second, a difference in the secondary eclipse depth of a hot Jupiter as measured by \cheops{} and \tess{} can be used to disambiguate the reflected light from the planet and its thermal emission.  Among the catalog of 1948 predicted \tess{} detections by \citet{Sullivan2015}, for only $\sim5$ planets will the expected difference in eclipse depth significantly exceed the \tess{} photometric error alone.  However, since the contribution of thermal emission to the transit depth is important in each of those cases, observations of these by \cheops{} should be a priority.

Third, the presence of an atmosphere around a planet can produce a difference in the transit depths and since atmospheres are more opaque/scattering at shorter wavelengths, the transit depth observed by \cheops{}, which is more sensitive to bluer light, will be slightly deeper than that observed by \tess{}. The effect will be negligible if the atmosphere has a haze/cloud of very small or large particles, or is clear but has significant water vapor which "flattens" the effect by Rayleigh scattering.  A larger difference can occur if the planet is both dry and haze/cloud free.  Since water is presumed to have accreted as ice in small planets, if the close-in planets detected by \tess{} lack water it is an indication that they accreted inside the "ice line" rather than migrating inwards from outside.  The largest differences occur for hazy atmospheres containing particles with a characteristic size of $\sim 0.05\mu$m.  If each \tess{}-detected planet possessed such an atmosphere, we estimate that $\sim 100$ would be detected by comparing \tess{} and \cheops{} photometry, depending on the atmospheres' heavy element abundance.

\begin{figure}
 \includegraphics[width=\columnwidth]{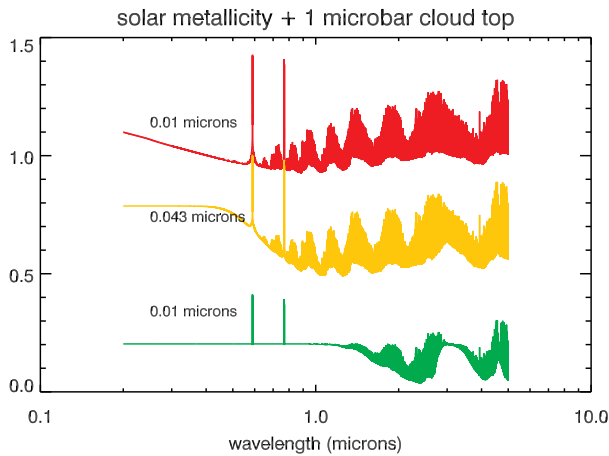}
 \caption{Three simulated, normalized transmission spectra of a GJ~1214b with a hazy solar-metallicity atmosphere.  The cloud top is at 1 $\mu$bar in each case, but the particle sizes differ.  Only the intermediate particle size (0.043$\mu$m) case produces a detectable difference in \cheops{}-\tess{} transit depths, but the haze does not significantly affect the strength of molecular bands at infrared wavelengths.}
 \label{fig.spec}
\end{figure}

The hazes detected by a comparison of \cheops{} and \tess{} photometry do not preclude investigation of the atmospheres at the longer wavelengths to be probed by \jwst{} and \ariel{}.  Figure \ref{fig.spec} plots three normalized model transit spectra, all for a GJ~1214b analog with a solar-metallicity atmosphere, \teq{}=1000K, and a 1 $\mu$bar cloud-top, but with three different particle sizes.  The intermediate  size particles (middle, yellow curve) produces the largest \cheops{}-\tess{} signal while the smaller and larger particle sizes are essentially undetectable.  The 0.043 $\mu$m haze particles have a negligible effect on the strength of the \water{} and other molecular bands in the infrared.  Therefore, detection of Rayleigh scattering by \cheops{}+\tess{} is a valid selection criterion for targets for \jwst{} or \ariel{}.  On the other hand, non-detection does not necessarily mean that the planets lack an atmosphere that could be observed in the infrared (e.g. very small haze particles, no haze at all, and/or very high mean molecular weight).  

Other astrophysical effects and systematic errors could produce differences in the transit depth as measured by the two observatories.   These include wavelength-dependent limb-darkening; at the shorter wavelengths to which \cheops{} is more sensitive than \tess{}, the limb of a star is darker relative to the center of the disk.  This will make a transit with a non-zero impact parameter appear shallower in \cheops{} data.  This will produce a {\it negative} \cheops{}-\tess{} transit depth difference.  A star spot that is crossed by the planet during the transit that is observed by \tess{} but not by \cheops (or vice verse) can cause a difference of either sign, but these events are often obvious since the spot crossing time is much shorter than the transit duration.  Non-occulted spots on the stellar disk will cause the transit signal to be deeper at bluer wavelengths, thus mimicking the effect of Rayleigh scattering in a planet's atmosphere \citep{Oshagh2014}.  This is a potentially serious problem for transit spectroscopy in general, but might be controlled by constraining the spot distribution by other observations.  Finally, there may be a systematic instrumental offset between \cheops and \tess{}, although this can be identified and removed if enough transiting systems are observed.

As shown in Fig. \ref{fig.bandpass} and discussed by \citet{Angerhausen2015}, the band-pass of \kep{}/K2 is similar to that of \cheops{} and in principle the same analyses could be performed using \tess{} and \kep{}/K2 data.  However, the \kep{} prime mission observed only $\approx 115$ sq deg. of the sky and only a handful of overlapping detections.  K2 is restricted to the ecliptic plane which the primary \tess{} mission will {\it not} observe.  Comparison of \tess{} observations with the {\it PLATO} mission \citep{Rauer2014}, scheduled for launch in 2025, will provide a more promising avenue for inter-comparison.  {\it PLATO} will observe several short-duration ($\sim 3$ month) and long-duration (up to 2 years) fields.  The response functions of the {\it PLATO} cameras are not yet determined, but all but two will operate without a filter and should have a total response function similar to that of \kep{} and \cheops{}.  Two high-cadence cameras will have distinct red and blue filters, which will offer color information for systems in a more limited part of the sky. 

\begin{figure}
 \includegraphics[width=\columnwidth]{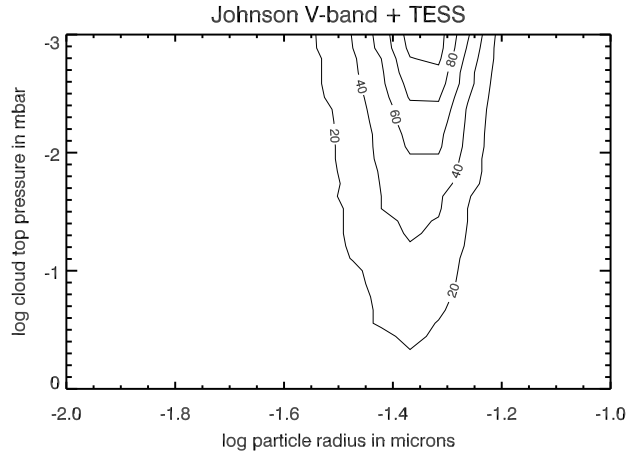}
 \caption{Numbers of $5\sigma$-detected hazy, solar-metallicity atmospheres (\teq = 1000K) around all simulated \tess{}-detected planets in the catalog of \citet{Sullivan2015}, based on the expected difference between transit depths measured by \tess{} and ground-based observations through a Johnson $V$ filter.}
 \label{fig.vband}
\end{figure}

Ground-based transit photometry can be performed through a pass-band that has little or no overlap with \tess{}, e.g. $V$ or $JHK$, allowing larger signals to be detected, albeit with less accuracy.  We calculated the fractional difference in transit depths between \tess{} and Johnson $V$-band for the scenario of a hazy atmosphere on a GJ~1214b-like planet.  We then calculated the number of possible detections among the \citet{Sullivan2015} catalog of simulated \tess{} planets.  We assumed $10^{-3}$ photometry per 1 minute observation and a reduction in noise with observation time $t$ as $t^{-0.3}$.  The use of -0.3 as the exponent rather than the -0.5 of uncorrelated (``white'') noise reflects the importance of correlated (``red'') noise in limiting ground-based observations \citep{Pont2006}.  These parameters produce a median photometric accuracy over the transit interval of 230 ppm, consistent with the best performance achieved on 10-meter class telescopes \citep{Nortmann2016}, but well short of the predicted median of 120 ppm for \tess{} observations \citet{Sullivan2015}.  Numbers of detected hazy, solar-composition atmospheres (\teq{} = 1000K) vs. particle size and cloud-top height are plotted in Fig. \ref{fig.vband}.  Even with this optimistic assumption for photometric accuracy, the numbers of detection are reduced by a factor of $\sim 2$ (maximum of 102) compared to \cheops{} observations.  This illustrates one limitation of ground-based compared to space-based observations:  others are weather and nocturnal observing windows.

An alternative is to fly an observatory capable of transit and eclipse multi-wavelength photometry or spectroscopy on a large number of exoplanets.  The proposed \ariel{} mission, which is one of three candidate ESA M3 missions for possible launch in 2026, would observe several hundred systems.  Selection of \tess{} targets based on a \cheops{}-\tess{} inter-comparison would improve the scientific yield of such a mission.  

\section*{Acknowledgments}

We thank A. Mann for providing the database of flux-calibrated spectra and A. Fortier for providing the engineering response functions for the \cheops{} camera.  EG thanks the University of Bern Center for Space and Habitability and the International Space Science Institute for support as a visiting scientist.  K.H. and D.K.  acknowledge partial financial support from the Center for Space and Habitability (CSH), the PlanetS National Center of Competence in Research (NCCR), the Swiss National Science Foundation and the MERAC Foundation. 

\clearpage

\appendix
\section{Detection of Wavelength-Dependent Phase of Light Curve Maxima}
\label{sec.phase}

In addition to detecting differences in the depth of secondary eclipses with different band-passes, it might be possible to detect differences in the phase of the maximum in the emitted plus reflected light curve of the planet with respect to the secondary eclipse.  These phase shifts are thought to be produced by equatorial jets on the rotating planet; heat is advected from the day side over the eastern terminator to the night side, causing the maximum in the thermal signal to lead the secondary effects, and clouds that form on the cooler night side being advected over the western terminator to the day side, causing the maximum in the reflected signal to lag the secondary eclipse \citep[e.g.,]{Shporer2015}.   For hot Jupiters where the observed signal is a mixture of reflected and emitted light, these two effects act against each other.  However, inspection of the phase curves in two pass-bands which have different relative contributions of thermal and reflected signal, the phase shift will change.  

Approximating the normalized phase curve in pass-band $A$ as the sum of two (reflected and thermal) sinusoidal signal shifted by $\theta$,
\begin{equation}
s(\theta) =  cos(\phi) + r cos(\phi + \theta), 
\end{equation}
where $r$ is the relative contribution of the thermal signal to a particular pass-band.  This function peaks at 
\begin{equation}
\phi = - \arctan \frac{r_A \sin \theta}{1 + r_A \cos \theta} 
\end{equation}
The relative shift $\delta \phi$ due to a small difference $\Delta r$ between two pass-bands is approximately:
\begin{equation}
\delta \phi \approx \frac{\delta r \sin \theta}{1 + 2r \cos \theta + r^2}
\end{equation}
Assuming the values for the Kepler-43 system and $\theta \sim 100$ deg., based on \citet{Angerhausen2015}, then $r$ is 0.31 and 0.09 for \tess{} and \cheops{}, respectively, and the phase difference $\phi \approx 13$~deg, which might be detectable; fits with \kep{} data typically have errors of a few degrees \citep{Angerhausen2015}. 

\end{document}